# Improving Supply Chain Coordination by Linking Dynamic Procurement Decision to Multi-Agent System

Yee Ming Chen

**Abstract**—The Internet has changed the way business is conducted in many ways. For example, in the field of procurement, the possibility to directly interact with a trading partner has given rise to new mechanisms in the supply chain management. One such interactive dynamic procurement, which lets both buyer and seller software agents bid by potential buyer agents instead of static procurement by vendors. Dynamic procurement decision could provide the buying and selling channel to buyer, to avoid occurring condition that seller could not deliver on the contract promise. Using NYOP(Name Your Own Price) to be the core of dynamic procurement negotiation algorithm sets up multi-agent dynamic supply chain system, to present the DSINs(Dynamic Supply Chain Information Networks) by JADE, and to present the dynamic supply chain logistic simulation by eM-Plant. Finally, evaluating supply chain performance with supply chain performance metrics (such as bullwhip, fill rate), to be the reference of enterprise making deciding in the future.

**Index Terms**—Multi-Agent System、Bullwhip effect、Fill rate、NYOP(Name Your Own Price)

————————— ◆ —————————

## 1 INTRODUCTION

Nowadays agent technology is utilized in many industrial applications such as production planning, collaborative engineering, and more recently supply chain management(SCM). One of the most difficult but critical issues in SCM is to improve the efficiency of supply chains from the perspective of the whole supply chain, not individual companies. More specifically, the formation, optimisation and minimizing the bullwhip effect(BWE) (the magnification of demand fluctuations as orders move up the supply chain.) [3] are considered critical issues for efficient supply chain management [14]. Current approaches for reducing the BWE can be distinguished in (1) information sharing (2) collaborative planning. Although it is well-known that sharing information among supply chain members can lead to improved efficiency, information sharing is not always possible, often because of the limitations in information systems[1][12]. Collaborating production and orders of supply chain members by a single company or a decision-making unit is in many cases infeasible, because supply chain members are usually independent companies. Considering the limitations of information sharing and collaborative planning, this paper proposes an agent-based Dynamic Supply Chain Information Networks (DSINs) in which each participant decides about procurement and production based on local information only. The interaction with other participants adopt a special form of Name Your Own Price(NYOP). The NYOP channel, exemplified by Priceline, is a popular online alternative to other, more traditional channels, through which service providers such as airlines, hotels, and car rental companies offer their products to customers. This paper is motivated by the importance of developing an understanding of the market implications of this pricing and distribution model. Therefore,NYOP plays a dominating and in many cases an exclusive role for coordinating supply and demand. The objective of our work is to design a dynamic procurement mechanism in the multi-agent system. Each SC participant is represented by an autonomous software agents that negotiation for coordinating supply and demand to reduce the BWE. The remainder of our paper is structured as follows: In section 2, we review existing work. Then we describe a dynamic procurement scheme that integrates concepts of NYOP. In section 4, we evaluate our proposal by conducting a simulation study using a multi-agent system(Dynamic Supply Chain Information Networks; DSINs). Section 5 discusses the pros and cons of DSINs. Finally, section 6 draws conclusions, and points out avenues of future research.

## 2. RELATED WORK

The related work to ours can be grouped into three major areas: using multiagent technology and information sharing in supply chains, countering the BWE, and adopting dynamic procurement for coordination in supply chains. Each area is briefly discussed below.

(1) Multi-agent technology and sharing information
Coordination plays a pivotal role in successful design and implementation of supply chains, especially for those that are formed by independent and autonomous companies[2]. In SCM, multiagent technology is at the brink of being integrated into real-world applications. An example

————————————————

• *Yee Ming Chen is with the Department of Industrial Engineering and Management, Yuan Ze University, Taoyuan, Taiwan.*

is MASCOT, which is a reconfigurable, multilevel, agent-based architecture for planning and scheduling, aimed at improving SC agility [13]. Chen and Wei [5] analyzed the effects of negotiation-based information sharing in a distributed make-to-order manufacturing supply chain in a multi-period, multi-product types environment, which is modeled as a multi-agent system. Verdicchio and Colombetti [6] also stress that information sharing as a critical factor for successful business process management. Supply chain decisions are improved with access to global information. However, supply chain partners are frequently hesitant to provide full access to all the information within an enterprise [10]. A mechanism to make decisions based on global information without complete access to that information is required for improved supply chain decision making. Cigoloni and Marco [2] carried out a study of the Collaborative Planning Forecasting and Replenishment (CPFR) process for trading partners (belonging to the same supply chain) who are willing to collaborate in exchanging sales and order forecasts. The two major out come of the literature survey is that information sharing is most important requirement of efficient supply chain and multi agent modelling is most suitable for designing of supply chains.

(2) Bullwhip effect
The term Bullwhip Effect was coined by Procter & Gamble management who noticed an amplification of information distortion as order information travelled up the supply chain [3]. The Bullwhip Effect (or Whiplash Effect) is an observed phenomenon in forecast-driven distribution channels. To address the Bullwhip Effect, many techniques are employed to manage various supply chain processes, such as order information sharing, demand forecasting, inventory management, and shipment scheduling [3]. Factors contributing to the Bullwhip Effect: forecast errors, overreaction to backlogs, lead time (of information – orders and of material) variability, no communication and no coordination up and down the supply chain, delay times for information and material flow, batch ordering (larger orders result in more variance), rationing and shortage gaming, price fluctuations, product promotions, free return policies, inflated orders. Bullwhip Effect is also attributed to the separate ownership of different stages of the supply chain. Each stage in such a structured supply chain tries to amplify the profit of the respective stages, thereby decreasing the overall profitability of the supply chain [13].

(3) Dynamic procurement
Dynamic procurement in a capacitated supply chain faces uncertain demand. Burnetas and Gilbert [1] study reservation or forward purchase decisions by providing the buyer a second means of capacity procurement after the demand uncertainty is revealed [7]. Fredergruen and Heching [6] study a model similar to ours and show that dynamic pricing/procurement redistributes the power between the supply chain partners and allows the buyer to impact the wholesale prices. The Internet allows companies to interact with customers in ways that may have been prohibitively costly using traditional channels [10][11]. In particular, the Internet has facilitated the emergence of name-your-own-price (NYOP) auctions. In such auctions, as popularized by Priceline, buyers bid for a product or service [8]. If a bid exceeds the seller's concealed threshold price, the buyer receives the product at the price of her bid. This selling mechanism has been used predominantly in the SC. Several normative models have been proposed regarding the optimal bid sequence [9][13]. In this section, we describe procurement decisions in SC with local information. This research employs a stylized model to identify and understand key tradeoffs driving the decision by a service provider in the SC to employ an NYOP channel, assuming that such a channel is available.

## 3. DYNAMIC PROCUREMENT SCHEME

In this section, we describe dynamic procurement scheme with NYOP in SC. The adaptation of NYOP for dynamic procurement calls for specifications of the three steps :

Step 1: Buyer agent's bid: For calculating the bid of the buyer agent we refer to the demand function. A linear demand curve, thus an increasing price P causes a decreasing demand Q

$$Q_d = a - bP \qquad (1)$$

where a>0 and b>0, the demand function describes the bidding behavior of the buyer agent.

Step 2: Seller agent's minimum price: For calculating the minimum price of the supplier we refer to the supply function. The linear form is

$$Q_s = c + dP \qquad (2)$$

where c>0 and d>0 .

Step 3: Matching: By comparing bid price and minimum price for the requested quantity, the price been decided.

Step 1 and 2 require that equilibrium price $P^*$ and equilibrium quantity $Q^*$ as well as estimates for the price elasticity of demand $E_d$ and supply $E_s$ are available. Then we can determine the parameters $a$ and $b$ of the demand function as well as $c$ and $d$ of the supply function. Figure 1 shows the relationship between demand and supply as described.

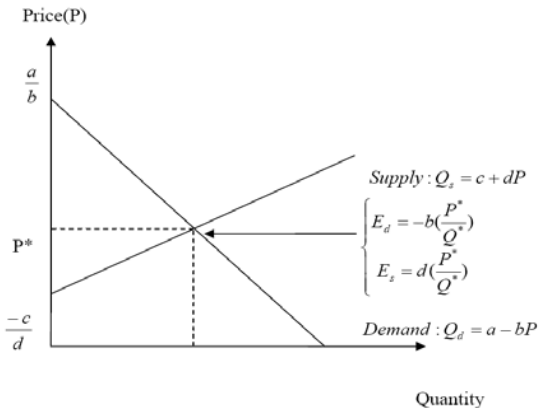

Figure 1   Demand and supply function

## 4, FORMATION OF BUYER-SUPPLIER AGENTSHIP WITH DYNAMIC PROCUREMENT SCHEME

Fig 1 shows an example supply chain of a Taiwam LCD panel assembly company which is used throughout this paper. LCD modules are widely used for display devices such as mobile phones, PDAs, and notebook PCs. In the supply chain, for a company like Cell Plant demand is distorted along the multiple paths from the markets to the company due to bullwhip effects and accumulation of erroneous estimations.

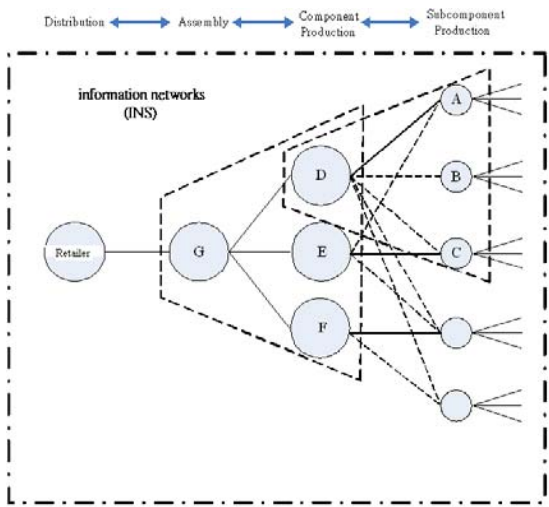

Figure 2 An example supply chain model of a LCD panel assembly manufacturing company

DSINs(Dynamic Supply Chain Information Networks) takes a simple but practical approach to address this problem. In an ADINS, agents autonomously form an information network by only local collaboration and information sharing. Fig 3 shows the architecture of a proposed DSINs. To increase the practicality of DSINs, all the interactions among participating agents in a supply chain are implemented with Java Agent Development Framework (JADE) which based on the specifications from FIPA ACL (Agent Communication Language) format [4].

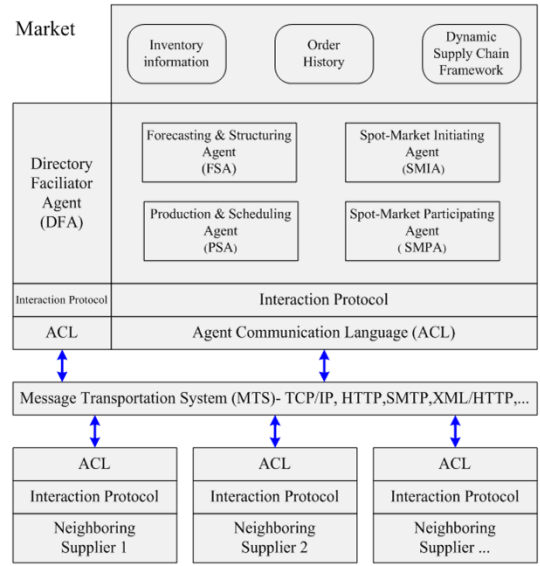

Figure 3  The architecture of DSINs

Using the network, they perform order and production planning in a synchronized way without any central controlling entities. For this, we assume that agents are able to observe market demands directly rather than relying on the possibly distorted demand figures that are received by their companies. By doing this, DSINs is able to reduce bullwhip effects and improve service quality such as fill rates.

In this multi-agent-based system, each SC participant is represented by an autonomous software agent. The central component of the JADE agent platform is the agent management system, which keeps supervisory control over access and use of the agent platform. Agents coordinate their ordering behavior by exchanging messages using the JADE message transport system. DSINs agents are generated using the JADE agent class and by implementing the appropriate agent control behavior on top of it. DSINs agents can send orders to other agents and receive goods. In this context, we distinguish two types of scenarios: Agents with/without NYOP scheme determine their orders according to the dynamic procurement negotiation algorithm (Figure 4.), and present the dynamic supply chain logistic simulation by eM-Plant.

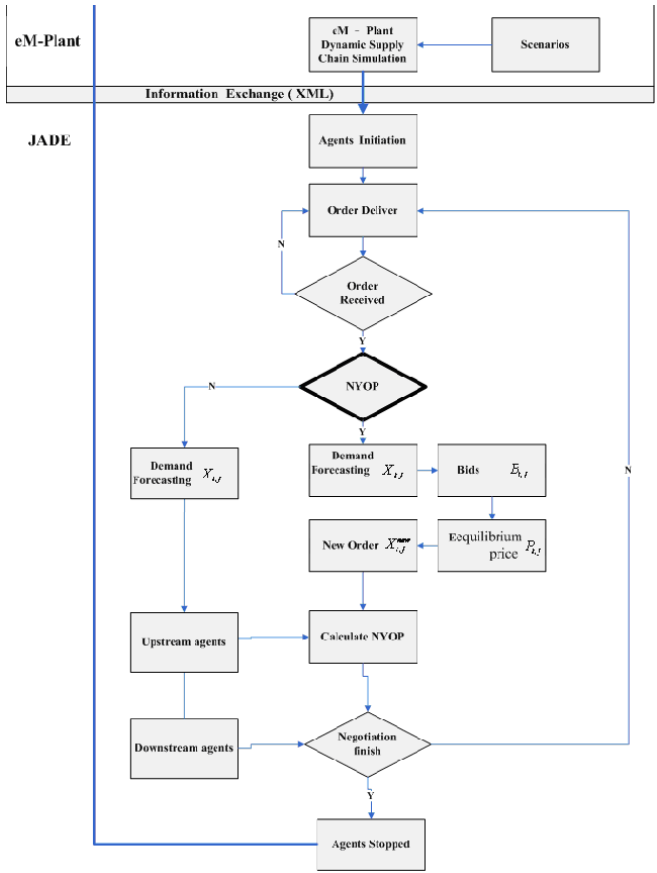

Figure 4.  Dynamic procurement negotiation algorithm

An agent interacts with other agents by placing orders for goods. An example of this interaction is shown in Figure 5: In each period T, an FSA agent ( forecasting & structuring agent)  observes the current inventory level and sends an order (calculated by algorithm ) to PSA (production & scheduling agent). After the order is placed, PSA agent observes and fills its demand for this period. Since our objective is to quantify the BWE, we must determine the variance of placed orders, relative to the variance of Demand. Figure 5 presents message exchanges captured in the experiment with the help of a JADE provided sniffer agent.

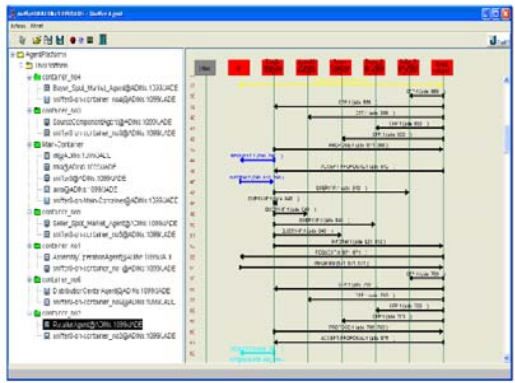

Figure 5  Screen captures showing our NYOP scheme in transaction.

## 5. EXPERIMENTS SETTING AND RESULTS

We defined a four-tier SC consisting of four agents (see also figure  2). Only the last agent ( $k = 4$ ) knows the final retailer demand  D . We set the following parameters:

(1) Final retailer demand  D : normally distributed with expected value $\mu$ =100 and standard deviation $\sigma$ =10 .
(2) The price elasticity of demand $E_d$ =-0.75 and supply $E_s$ =1.56.
(3) $T$  between 5 and 15 (time periods of the moving average forecast).
(4) $k$ between 1 and 4 (same lead time for all tiers)

Table 1  BWE for k=4  (Retailer)
(SD: standard deviation , T：number of periods)

| k=4 | Without NYOP | | With NYOP | | Change | |
|---|---|---|---|---|---|---|
| T | Mean | SD | Mean | SD | Mean | SD |
| 5 | 7.84 | 0.74 | 5.94 | 0.61 | 24.23% | 17.57% |
| 6 | 7.61 | 0.70 | 5.72 | 0.54 | 24.84% | 22.86% |
| 7 | 7.21 | 0.63 | 5.50 | 0.48 | 23.72% | 23.81% |
| 8 | 7.09 | 0.57 | 5.46 | 0.46 | 22.99% | 19.30% |
| 9 | 6.83 | 0.55 | 5.28 | 0.42 | 22.69% | 23.64% |
| 10 | 6.29 | 0.49 | 4.95 | 0.40 | 21.30% | 18.37% |
| 11 | 6.13 | 0.47 | 4.61 | 0.37 | 24.80% | 21.28% |
| 12 | 5.84 | 0.44 | 4.15 | 0.34 | 28.94% | 22.73% |
| 13 | 5.41 | 0.41 | 3.86 | 0.33 | 28.65% | 19.51% |
| 14 | 5.38 | 0.40 | 3.58 | 0.30 | 33.46% | 25.00% |
| 15 | 5.30 | 0.37 | 3.19 | 0.26 | 39.81% | 29.73% |
| | | | | Mean of Change | 26.86% | 22.16% |

Table 2  BWE for k=3 (Wholeseller)

| k=3 | Without NYOP | | With NYOP | | Change | |
|---|---|---|---|---|---|---|
| T | Mean | SD | Mean | SD | Mean | SD |
| 5 | 9.35 | 0.83 | 6.87 | 0.78 | 26.52% | 6.02% |
| 6 | 8.87 | 0.79 | 6.65 | 0.71 | 25.03% | 10.13% |
| 7 | 8.64 | 0.75 | 6.43 | 0.67 | 25.58% | 10.67% |
| 8 | 8.24 | 0.73 | 6.39 | 0.61 | 22.45% | 16.44% |
| 9 | 8.12 | 0.72 | 6.21 | 0.57 | 23.52% | 20.83% |
| 10 | 7.86 | 0.67 | 5.98 | 0.57 | 23.92% | 14.93% |
| 11 | 7.32 | 0.65 | 5.64 | 0.52 | 22.95% | 20.00% |
| 12 | 7.14 | 0.62 | 5.24 | 0.47 | 26.61% | 24.19% |
| 13 | 6.84 | 0.58 | 5.01 | 0.43 | 26.75% | 25.86% |
| 14 | 6.65 | 0.57 | 4.79 | 0.42 | 27.97% | 26.32% |
| 15 | 6.21 | 0.53 | 4.46 | 0.43 | 28.18% | 18.87% |
| | | | | Mean of Change | 25.41% | 17.66% |

Table 3  BWE for k=2 (Manufactuer)

| k=2 | Without NYOP | | With NYOP | | Change | |
|---|---|---|---|---|---|---|
| T | Mean | SD | Mean | SD | Mean | SD |
| 5 | 13.46 | 1.54 | 11.21 | 1.43 | 16.72% | 7.14% |
| 6 | 12.63 | 1.51 | 10.77 | 1.41 | 14.73% | 6.62% |
| 7 | 10.85 | 1.43 | 9.53 | 1.28 | 12.17% | 10.49% |
| 8 | 9.95 | 1.39 | 8.12 | 1.25 | 18.39% | 10.07% |
| 9 | 9.67 | 1.34 | 7.94 | 1.21 | 17.89% | 9.70% |
| 10 | 8.79 | 1.27 | 7.71 | 1.13 | 12.29% | 11.02% |
| 11 | 8.34 | 1.24 | 7.37 | 1.10 | 11.63% | 11.29% |
| 12 | 7.87 | 1.21 | 6.97 | 1.05 | 11.44% | 13.22% |
| 13 | 7.59 | 1.2 | 6.74 | 1.05 | 11.20% | 12.50% |
| 14 | 7.46 | 1.18 | 6.52 | 1.04 | 12.60% | 11.86% |
| 15 | 7.24 | 1.15 | 6.19 | 1.02 | 14.50% | 11.30% |
| | | | | Mean of Change | 13.96% | 10.48% |

Table 4  BWE for k=1 (Supplier)

| k=1 | Without NYOP | | With NYOP | | Change | |
|---|---|---|---|---|---|---|
| T | Mean | SD | Mean | SD | Mean | SD |
| 5 | 16.48 | 1.68 | 14.32 | 1.59 | 13.11% | 5.36% |
| 6 | 15.65 | 1.65 | 13.88 | 1.57 | 11.31% | 4.85% |
| 7 | 13.87 | 1.51 | 12.64 | 1.45 | 8.87% | 3.97% |
| 8 | 12.97 | 1.47 | 11.23 | 1.38 | 13.42% | 6.12% |
| 9 | 12.69 | 1.42 | 11.01 | 1.24 | 13.24% | 12.68% |
| 10 | 10.32 | 1.35 | 9.89 | 1.15 | 4.17% | 14.81% |
| 11 | 9.47 | 1.31 | 9.12 | 1.07 | 3.70% | 18.32% |
| 12 | 8.12 | 1.25 | 7.56 | 0.96 | 6.90% | 23.20% |
| 13 | 7.56 | 1.24 | 7.19 | 0.89 | 4.89% | 28.23% |
| 14 | 7.43 | 1.22 | 7.02 | 0.83 | 5.52% | 31.97% |
| 15 | 7.21 | 1.19 | 6.98 | 0.81 | 3.19% | 31.93% |
| | | | | Mean of Change | 8.03% | 16.49% |

In the first set of experiments, we determined the BWE under variation of $T$. From Table 1 to Table 4 present the data for the comparison of the conventional procurement decision and the NYOP scheme yields a reduction in the order BWE of 8.03 % to 26.86 %. In general, the BWE decreases with an increasing T (see also figure 6).

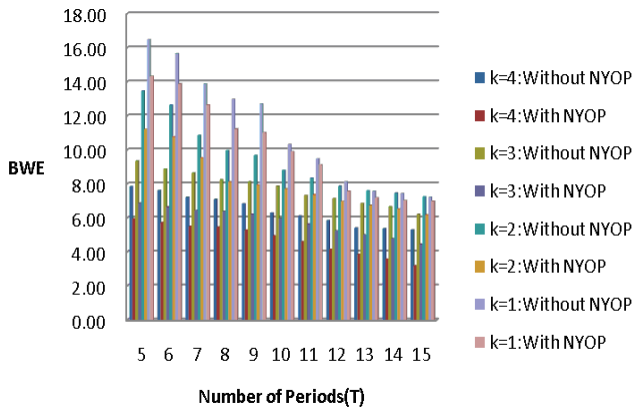

Figure 6   BWE and variation T

In the second set of experiments, we determined the Fill rate (FR) under variation of $T$.

$$FR_k = 1 - G_u(z)\sqrt{L_k}\sqrt{Var(Q_{k-1})}/\mu_k \quad k=1,2,3,4 \quad (3)$$

Where $G_u(z)$ is the standardized loss function.

Table 5  Fill Rate for k=4  (Retailer)
(SD: standard deviation , T：number of periods)

| k=4 | Without NYOP | | With NYOP | | Change | |
|---|---|---|---|---|---|---|
| T | Mean | SD | Mean | SD | Mean | SD |
| 5 | 0.61 | 0.53 | 0.62 | 0.54 | 1.61% | 1.85% |
| 6 | 0.62 | 0.54 | 0.64 | 0.56 | 3.13% | 3.57% |
| 7 | 0.64 | 0.56 | 0.65 | 0.57 | 1.54% | 1.75% |
| 8 | 0.68 | 0.58 | 0.69 | 0.58 | 1.45% | 0.00% |
| 9 | 0.71 | 0.59 | 0.72 | 0.60 | 1.39% | 1.67% |
| 10 | 0.72 | 0.61 | 0.77 | 0.62 | 6.49% | 1.61% |
| 11 | 0.75 | 0.62 | 0.78 | 0.63 | 3.85% | 1.59% |
| 12 | 0.77 | 0.62 | 0.82 | 0.65 | 6.10% | 4.62% |
| 13 | 0.78 | 0.63 | 0.83 | 0.65 | 6.02% | 3.08% |
| 14 | 0.78 | 0.64 | 0.84 | 0.66 | 7.14% | 3.03% |
| 15 | 0.79 | 0.66 | 0.81 | 0.67 | 2.47% | 1.49% |
| | | | | Mean of Change | 3.74% | 2.21% |

Table 6  Fill Rate for k=3  (Wholeseller)

| k=3 | Without NYOP | | With NYOP | | Change | |
|---|---|---|---|---|---|---|
| T | Mean | SD | Mean | SD | Mean | SD |
| 5 | 0.63 | 0.54 | 0.63 | 0.54 | 0.00% | 0.00% |
| 6 | 0.63 | 0.55 | 0.65 | 0.56 | 3.08% | 1.79% |
| 7 | 0.65 | 0.55 | 0.67 | 0.57 | 2.99% | 3.51% |
| 8 | 0.66 | 0.56 | 0.68 | 0.58 | 2.94% | 3.45% |
| 9 | 0.68 | 0.58 | 0.70 | 0.58 | 2.86% | 0.00% |
| 10 | 0.72 | 0.59 | 0.71 | 0.61 | -1.41% | 3.28% |
| 11 | 0.75 | 0.61 | 0.77 | 0.62 | 2.60% | 1.61% |
| 12 | 0.76 | 0.64 | 0.78 | 0.65 | 2.56% | 1.54% |
| 13 | 0.79 | 0.64 | 0.82 | 0.67 | 3.66% | 4.48% |
| 14 | 0.81 | 0.65 | 0.83 | 0.66 | 2.41% | 1.52% |
| 15 | 0.81 | 0.67 | 0.84 | 0.67 | 3.57% | 0.00% |
| | | | | Mean of Change | 2.30% | 1.92% |

Table 7  Fill Rate for k=2  (Manufactuer)

| k=2 | Without NYOP | | With NYOP | | Change | |
|---|---|---|---|---|---|---|
| T | Mean | SD | Mean | SD | Mean | SD |
| 5 | 0.63 | 0.53 | 0.63 | 0.54 | 0.00% | 1.85% |
| 6 | 0.64 | 0.54 | 0.65 | 0.56 | 1.54% | 3.57% |
| 7 | 0.65 | 0.56 | 0.67 | 0.57 | 2.99% | 1.75% |
| 8 | 0.66 | 0.57 | 0.68 | 0.57 | 2.94% | 0.00% |
| 9 | 0.69 | 0.58 | 0.71 | 0.57 | 2.82% | -1.75% |
| 10 | 0.70 | 0.60 | 0.72 | 0.58 | 2.78% | -3.45% |
| 11 | 0.72 | 0.61 | 0.74 | 0.62 | 2.70% | 1.61% |
| 12 | 0.75 | 0.62 | 0.77 | 0.64 | 2.60% | 3.13% |
| 13 | 0.78 | 0.62 | 0.79 | 0.63 | 1.27% | 1.59% |
| 14 | 0.79 | 0.63 | 0.80 | 0.64 | 1.25% | 1.56% |
| 15 | 0.81 | 0.65 | 0.81 | 0.65 | 0.00% | 0.00% |
| | | | | Mean of Change | 1.90% | 0.90% |

Table 8  Fill Rate for k=1 (Supplier)

| k=1 | Without NYOP | | With NYOP | | Change | |
|---|---|---|---|---|---|---|
| T | Mean | SD | Mean | SD | Mean | SD |
| 5 | 0.64 | 0.53 | 0.67 | 0.54 | 4.48% | 1.85% |
| 6 | 0.67 | 0.54 | 0.67 | 0.55 | 0.00% | 1.82% |
| 7 | 0.67 | 0.55 | 0.68 | 0.56 | 1.47% | 1.79% |
| 8 | 0.69 | 0.56 | 0.70 | 0.57 | 1.43% | 1.75% |
| 9 | 0.72 | 0.58 | 0.73 | 0.57 | 1.37% | -1.75% |
| 10 | 0.73 | 0.60 | 0.73 | 0.59 | 0.00% | -1.69% |
| 11 | 0.75 | 0.61 | 0.76 | 0.60 | 1.32% | -1.67% |
| 12 | 0.76 | 0.61 | 0.78 | 0.61 | 2.56% | 0.00% |
| 13 | 0.78 | 0.62 | 0.79 | 0.63 | 1.27% | 1.59% |
| 14 | 0.81 | 0.63 | 0.83 | 0.63 | 2.41% | 0.00% |
| 15 | 0.83 | 0.63 | 0.85 | 0.64 | 2.35% | 1.56% |
| | | | | Mean of Change | 1.70% | 0.48% |

In Table 5 ~ 8, the fill rate improve from 1.70 % to 3.74 % with NYOP scheme against the conventional procurement decision in four tier supply partners. The FR increases with an increasing T (figure 7).

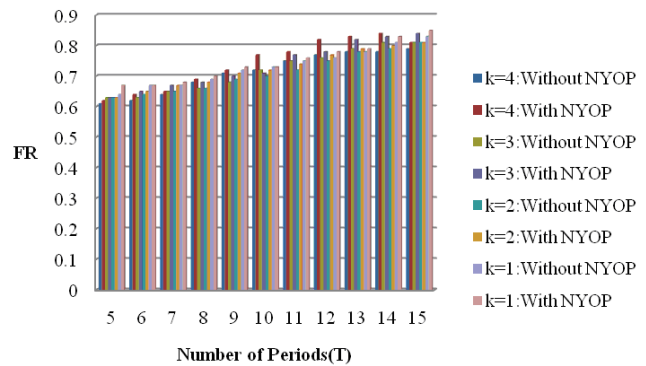

Figure 7   Fill rate and variation T

## 6. CONCLUSIONS AND FUTURE WORK

We aimed at reducing the bullwhip effect and improve fill rate in multi-tier SCs by means of NYOP scheme. In particular, the multiagent based behaviour model for each agent was developed to clarify the states, transitions, and communication requirements of agents and also to facilitate the derivation of concrete procedures for agent behaviours that can be used for actual development of agent systems. In order to show the practical feasibility of the approach, the conversations among agents were also modelled with FIPA's standard interaction protocols and messages and a prototype DSINs system was constructed using a FIPA-compliant agent platform JADE.

A major limitation is that we assumed linear supply and demand functions. This assumption is, however, coherent with basic concepts of market economics, we believe that this paper has contributed to proving the ever-growing potential of agent technology for practical supply chain management where analytic or optimization results from multi-tier supply chains cannot be easily applied or global information sharing or central coordination is impossible. Further rsearch issues include the full implementation of DSINs, relaxing the assumptions on supply chains, and analysing the bidding strategies for the iterative relaxation Contract Net in the supply chain management.

**Yee Ming Chen** is a professor in the Department of Industrial Engineering and Management at Yuan Ze University, where he carries out basic and applied research in agent-based computing. His current research interests include soft computing, supply chain management, and pattern recognition.